\documentclass[twocolumn, prl, showpacs,superscriptaddress]{revtex4}
\usepackage{amssymb}
\usepackage{graphicx}
\usepackage{dcolumn}
\usepackage{bm}
\usepackage{amsmath}

\begin{document}

\title{Two Atoms in a Double Well: An Exact Solution}
\author{Yanxia Liu}
\affiliation{Institute of Theoretical Physics, Shanxi University, Taiyuan 030006, P. R.
China}
\author{Yunbo Zhang}
\email{ybzhang@sxu.edu.cn}
\affiliation{Institute of Theoretical Physics, Shanxi University, Taiyuan 030006, P. R.
China}

\begin{abstract}
We propose to experimentally realize an odd parity
eigenstate $\left\vert b\right\rangle $ of two atoms in the double well. The
occupation probability of this state shows evident
dependence on the interaction, distinct from the result of two-mode model
adopted in the Heidelberg experiment. The tunneling
dynamics of two atoms starting from the $NOON$ state
with infinite barrier height can be derived from 
the exactly solved model of $\delta $-barrier split double well 
based on a Bethe ansatz type hypothesis of the
wave functions.
We find that the single particle tunneling transfer the probabilities between
double occupancy and single occupancy of each well.
\end{abstract}

\maketitle

\textit{Introduction}: The deterministic preparation of few-particle systems
makes it possible to study the tunneling of a few atoms out of a metastable
state \cite{Zurn2012,Massimo,Zurn2013} or oscillation in double well \cite%
{Cheinet,Murmann}. The double well is a typical model for the observation of
Josephson oscillations in superconductor qubit \cite{Cooper,Simmonds} and
nonlinear self-trapping of Bose-Einstein condensates \cite{Albiez,Shin,Saba}.
However, most studies so far relied on the two-mode model valid for
sufficiently weak coupling \cite{Milburn,Smerzi} or numerical methods for
the dynamics \cite{Zollner,Murphy,Yin2008,Lv2010}.

Recently, the Heidelberg group has reported a new breakthrough in realizing
two ultra-cold $^{6}$Li fermonic atoms in an isolated double-well potential,
which constitutes the fundamental building block of the Hubbard model at
half filling \cite{Murmann}. In this experiment, the ground state $%
\left\vert a\right\rangle $ and the excited state $\left\vert c\right\rangle 
$ can be prepared with different interaction strength. The double occupancy
of the two states is suppressed for increasing repulsive interactions. They
did not bother to give the occupation statistics of the state $\left\vert
b\right\rangle $, which are constants in the two-mode approximation.
Practically tunneling of two atoms in a double well involves more single
particle states and the occupation statistics of the state $\left\vert
b\right\rangle $ changes with interaction strength. The spatial wave
functions of two fermions in a spin-singlet configuration $\left\vert
a\right\rangle $, $\left\vert b\right\rangle $ and $\left\vert
c\right\rangle $ are symmetric with respect to particle exchange, which is
identical to the case of two spinless bosons. On the other hand, the system
has space inversion symmetry expressed as $\Psi (x_{1},x_{2})=\pm \Psi
(-x_{1},-x_{2})$ and the eigenstates can be classified into even parity with
the sign "$+$" and odd parity with the sign "$-$". The states $\left\vert
a\right\rangle $ and $\left\vert c\right\rangle $ are even parity and the
state $\left\vert b\right\rangle $ is odd parity. The two-particle $NOON$
state $\left\vert b\right\rangle $ plays important role in several metrology
proposals, including atomic frequency measurements \cite{Bollinger},
interferometry \cite{Holland,Campos}, and matter-wave gyroscopes \cite%
{Dowling}. 
\begin{figure}[tbp]
\includegraphics[width=0.50\textwidth]{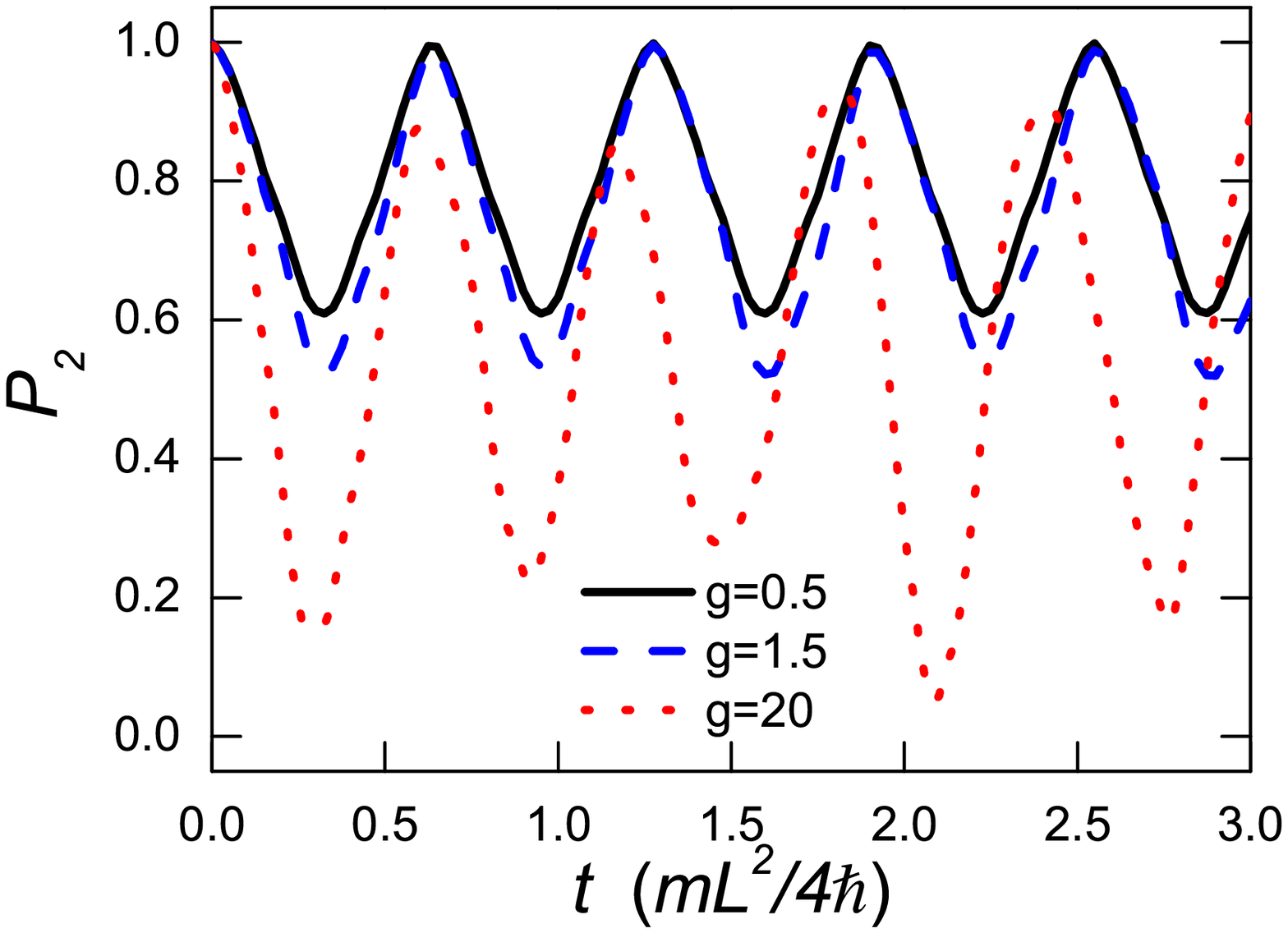} %
\includegraphics[width=0.50\textwidth]{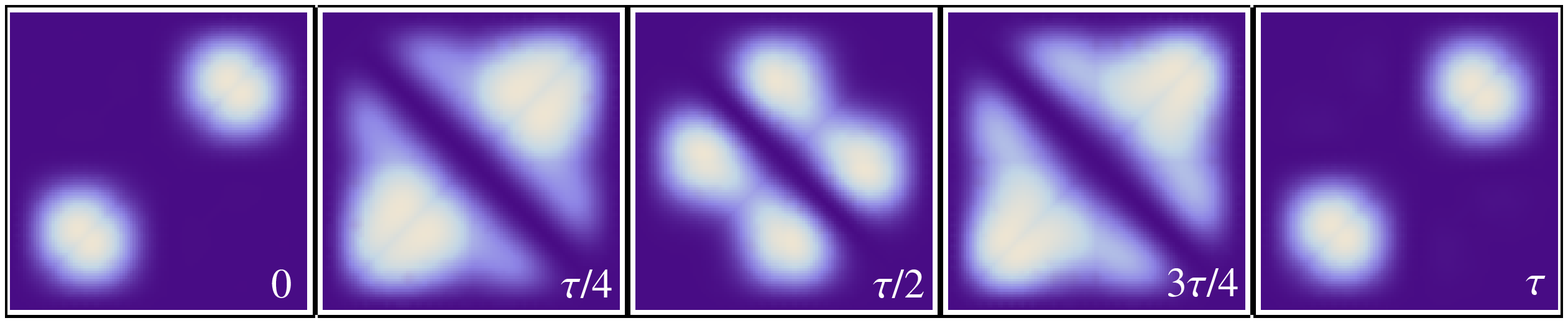} %
\includegraphics[width=0.50\textwidth]{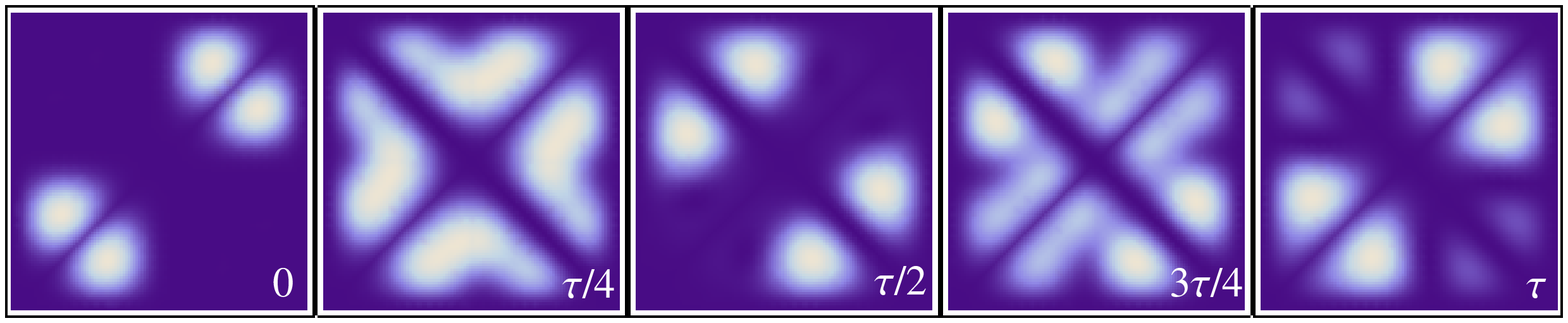}
\caption{(Color online). Top: Tunneling dynamics of the occupation
probability $P_{2}\left( t\right) $ of finding both atoms in the same well
for $g=0.5$ (black solid line), $g=1.5$ (blue dashed line) and $g=20$ (red
dotted line). The $\protect\delta$-barrier is abruptly lowered from a height 
$d=300$ to $0.5$ at time $t=0$. Bottom: The two-body density functions $%
\protect\rho \left( x_{1},x_{2},t\right) $ at different times $t$, for $%
g=1.5 $ and $20$, respectively.}
\label{fig1}
\end{figure}

Motivated by the above experiment, in this Letter we choose the state $%
\left\vert b\right\rangle $ as the initial state in double well and
investigate the tunneling dynamics by an exactly solved model. There are a
handful of exactly solved models in one dimension (1D), for instance,
Lieb-Liniger model \cite{Lieb}, Gaudin-Yang model \cite{Gaudin,Yang}, etc 
\cite{MTakahashi}. Recent experiments on ultracold bosonic and fermionic
atoms confined to 1D have provided a better understanding of the quantum
statistical and dynamical effects in quantum many-body systems \cite{GuanRMP}%
. We give a new formulation on the exact solution of a $\delta $-split
double well. Based on the analytically derived Bethe-ansatz-like equations
and the Yang-Baxter-like relations for scattering matrices, we propose a
scheme to achieve accurate tunneling dynamics for two atoms in the $NOON$
state of a double well as shown in Fig. \ref{fig1}.

\textit{System}: We start with the model of $N$ bosonic atoms of mass $m$
confined in a 1D tube of length $L$, split by a $\delta $-function potential
barrier at the center. The atoms interact through $s$-wave scattering and
the Hamiltonian of the system can be written as%
\begin{equation}
H=\sum_{i=1}^{N}\left[ -\frac{\hbar ^{2}}{2m}\frac{\partial ^{2}}{\partial
x_{i}^{2}}+d\delta \left( x_{i}\right) +V(x_{i})\right] +g\sum_{i<j}^{N}%
\delta \left( x_{i}-x_{j}\right) ,  \label{N1}
\end{equation}%
where $g$ is the interaction strength which can be tuned via magnetic
Feshbach resonance and confinement induced resonance from $-\infty $ to $%
+\infty $, and $d>0$ is the height of the $\delta $-function barrier. The
square well $V(x)$ with infinite depth is equivalent to the open boundary
condition $\Psi \left( x_{i}=\pm L/2\right) =0,i=1,2,\cdots ,N$, for the
eigenfunction $\Psi (x_{1},x_{2},\cdots ,x_{N})$ of Hamiltonian $H$. The
system in the absence of the $\delta $-barrier ($d=0$) reduces to the
Lieb-Liniger model with open boundary conditions \cite{Hao2006}, while in
the non-interacting case $g=0$ the problem turns out to be a single particle
in split well. The system permits only solutions with odd parity as the
situation when two particles simultaneously scatter on the $d$-barrier will
not be considered here, just as the three-body interaction is omitted in
Lieb-Lininger model. The addition of $\delta $-function barrier makes the
system highly nontrivial and the ground-state properties of few-particle
system in the split hard-wall potential have been examined in detail \cite%
{Yin2008,Lv2010} by means of the Bose-Fermi mapping in the Tonks-Giradeau
limit and exact diagonalization method for finite interaction. Here we try
to give an analytical solution in terms of Bethe's hypothesis and suitable
connection conditions for wave functions.

The wave function is taken as the Bethe ansatz (BA) type 
\begin{eqnarray}
&&\Psi (x_{1},x_{2},\cdots ,x_{N})  \notag \\
&=&\sum_{n=0}^{N}\sum_{Q,P}\sum_{\epsilon _{P}=\pm 1}\theta \left(
x_{Q_{1}}<x_{Q_{2}}<\cdots <x_{Q_{N}}\right)  \notag \\
&&\times A_{n}\left( Q,\epsilon _{P}P\right) \exp \left(
i\sum_{l=1}^{N}\epsilon _{P_{l}}k_{P_{l}}x_{Q_{l}}\right) ,  \label{N2}
\end{eqnarray}%
where $Q=\left( Q_{1},Q_{2},\cdots ,Q_{N}\right) $ and $P=\left(
P_{1},P_{2},\cdots ,P_{N}\right) $ are two arrangements of $\left(
1,2,\cdots ,N\right) $. For the arrangement of the coordinates, we assume
that $n$ coordinates are less than $0$, and the rest are larger than zero
such that $\left( x_{Q_{1}}<x_{Q_{2}}<\cdots <x_{Q_{n}}<0<x_{Q_{n+1}}\cdots
<x_{Q_{N}}\right) $. That is, $n$ atoms are on the left side of the $\delta $%
-barrier and $N-n$ of them are on the right side. We label such a block by a
single index $n$, which is a square for $N=2$, a cubic for $N=3$, and so on.
It constitutes of $N!$ local regions in the sense of Lieb-Lininger model.
The summation over $n$ in Eq. $\left( \ref{N2}\right) $ includes all
possible $N+1$ blocks. $\epsilon _{P}$ indicates that the particles move
toward the right $\left( \epsilon _{P}=+1\right) $ or the left $\left(
\epsilon _{P}=-1\right) $ and $\theta $ is the step function. This system is
featured by the fact that both interaction of two particles and the
potential barrier are $\delta $-functions. Generally each $\delta $-function
corresponds to a jump condition of the first derivative of wave function,
which is used to connect adjointing regions. In presence of the potential
barrier the momentum set $\{k\} $ satisfy a series of transcendental
equations, which play the role of BA equations \cite{Liu} 
\begin{equation}
ik_{P_{n}}\left( \frac{1-R_{n-1}\left( P_{n}\right) }{1+R_{n-1}\left(
P_{n}\right) }-\frac{1-R_{n}\left( P_{n}\right) }{1+R_{n}\left( P_{n}\right) 
}\right) =\frac{2m}{\hbar ^{2}}d,  \label{N21}
\end{equation}%
with $n$, $P_{n}=1,2,\cdots ,N$. Here we define the reflection matrix $%
R_{n-1}\left( P_{n}\right) $ as the ratio between coefficients $A$ after and
before the reverse of $P_{n}$ in block $n-1$, which reverses the sign of the
momentum of the $n$-th particle 
\begin{eqnarray}
R_{n-1}\left( P_{n}\right) &=&\frac{A_{n-1}\left( Q,\cdots ,-P_{n},\cdots
\right) }{A_{n-1}\left( Q,\cdots ,P_{n},\cdots \right) }  \notag \\
&=&-e^{ik_{P_{n}}L}\prod\limits_{i=1}^{N-n}\frac{k_{P_{n+i}}-k_{P_{n}}+i%
\frac{m}{\hbar ^{2}}g}{k_{P_{n+i}}-k_{P_{n}}-i\frac{m}{\hbar ^{2}}g}  \notag
\\
&&\times \frac{k_{P_{n+i}}+k_{P_{n}}-i\frac{m}{\hbar ^{2}}g}{%
k_{P_{n+i}}+k_{P_{n}}+i\frac{m}{\hbar ^{2}}g}.  \label{N17}
\end{eqnarray}%
In a similar way $R_{n}\left( P_{n}\right) $ inverses the momentum of the $n$%
-th particle in the block $n$ 
\begin{eqnarray}
R_{n}\left( P_{n}\right) &=&\frac{A_{n}\left( Q,\cdots ,-P_{n},\cdots
\right) }{A_{n}\left( Q,\cdots ,P_{n},\cdots \right) }  \notag \\
&=&-e^{-ik_{P_{n}}L}\prod\limits_{i=1}^{n-1}\frac{k_{P_{n}}-k_{P_{n-i}}+i%
\frac{m}{\hbar ^{2}}g}{k_{P_{n}}-k_{P_{n-i}}-i\frac{m}{\hbar ^{2}}g}  \notag
\\
&&\times \frac{k_{P_{n}}+k_{P_{n-i}}+i\frac{m}{\hbar ^{2}}g}{%
k_{P_{n}}+k_{P_{n-i}}-i\frac{m}{\hbar ^{2}}g}.  \label{N18}
\end{eqnarray}%
We note that the system is essentially the same when the number of atoms in
the left well equals $n$ or $N-n$. Due to this symmetry, there are
altogether $\frac{1}{2}N\sum\limits_{j=0}^{N-1}C_{N-1}^{j}=2^{N-2}N$ such
BA-type equations. For $d=0$, Eq. (\ref{N21}) is known as BA equation of the
open boundary condition. When $g=0$, Eq. (\ref{N21}) recovers the single
particle result $\tan \left( -k_{P_{n}}L/2+\hbar ^{2}k_{P_{n}}/md\right) =0$.

In the scattering situation here the atoms preserve their momentum while
changing their quantum internal states. The self-closed property of the
system requires that the order of the collisions does not affect the final
outcome. We find the Yang-Baxter equations (YBE)%
\begin{eqnarray}
&&S_{d,P_{n}}\left( n+1\right) S_{d,P_{n+1}}\left( n\right) S_{P_{n},P_{n+1}}
\notag \\
&=&S_{P_{n},P_{n+1}}S_{d,P_{n+1}}\left( n+1\right) S_{d,P_{n}}\left(
n\right) ,  \label{N22}
\end{eqnarray}%
which relies on the arrangement $P$ and describes two equivalent processes
from $\left( d,P_{n},P_{n+1}\right) $ to $\left( P_{n+1},P_{n},d\right) $.
Often a solution of YBE is referred to as an scattering matrix. Here the
scattering matrix between the $d$-barrier and the $n$-th particle is
calculated as 
\begin{eqnarray}
S_{d,P_{n}}\left( n\right) &=&\frac{A_{n}\left( Q,\cdots ,P_{n},\cdots
\right) }{A_{n-1}\left( Q,\cdots ,P_{n},\cdots \right) }  \notag \\
&=&\frac{1+R_{n-1}\left( P_{n}\right) }{1+R_{n}\left( P_{n}\right) },
\label{N19}
\end{eqnarray}%
which brings the system from block $n-1$ to $n$. The inverse scattering
process is defined as $S_{P_{n},d}\left( n\right) =1/S_{d,P_{n}}\left(
n\right) $, that is, the particle jumps from block $n$ back to $n-1$. The
index $n$ in the parentheses of $S_{d,P_{n}}\left( n\right) $ indicates the
position of $P_{n}$ in the arrangement $P$. The L.H.S of Eq. (\ref{N22})
shows that firstly two neighboring particles scatter with each other in
block $n-1$, than scatter with the $d$-barrier in turn. The R.H.S of Eq. (%
\ref{N22}) shows that they scatter with the $d$-barrier first, then exchange
their positions on the left of the $d$-barrier. The YBE is the basic
algebraic constituent in the quantum inverse method and have been found in
many model. The Eq. (\ref{N22}) is different from the regular YBE because
there exist two distinct forms of scattering matrices \cite{Liu}. Collisions
of two adjacent particles in different blocks satisfy the same scattering
matrix, so Eq. (\ref{N22}) is reduced to 
\begin{equation}
S_{d,P_{n}}\left( n+1\right) S_{d,P_{n+1}}\left( n\right)
=S_{d,P_{n+1}}\left( n+1\right) S_{d,P_{n}}\left( n\right).  \label{N25}
\end{equation}%
The same equation holds for blocks $n-1$ and $N-n+1$ and in general there
are $\frac{1}{4}N\left( N-1\right)
\sum\limits_{j=0}^{N-2}C_{N-2}^{j}=N\left( N-1\right) 2^{N-4}$ different
equations for $N>2$. When $N=2$, there is only one such equation. The
quasi-momentum $k_{i}$ can be obtained by solving the sets of equations of $%
\left( \ref{N21}\right) $ and $\left( \ref{N25}\right) $ and the energy
eigenvalue is $E=\sum_{i=1}^{N}\frac{\hbar ^{2}k_{i}^{2}}{2m}$. 
\begin{figure}[tbp]
\includegraphics[width=0.45\textwidth]{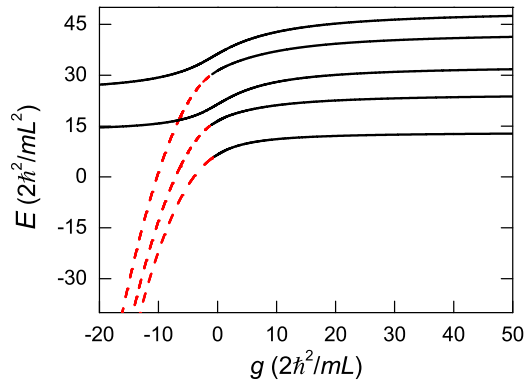}
\caption{(color online). Energy spectrum of two atoms in a double well. Five
lowest eigenstates with odd parity in 1D split hard-wall potential are shown
as a function of the interaction strength $g$ for barrier height $d=0.5$.
The solid curves (black) show the eigenenergies of the bound states for
atoms, the quasi-momentums of which are real numbers. The dashed curves
(red) show those of the molecule states and the corresponding
quasi-momentums are two conjugate complex numbers.}
\label{fig2}
\end{figure}

\textit{Tunneling dynamics in double-well:} We study the tunneling dynamics
of two atoms in the double-well. Two interacting atoms are initially
prepared in the $NOON$ state $\left\vert b\right\rangle $ of the double well
with an infinite barrier height (numerically we set $d=300$). Then the
barrier is abruptly lowered to a fixed height $d=0.5$, which allows the
atoms to tunnel between wells. Let the system evolve for different durations 
$t$ and the resulting dynamics is observed by quickly increasing the barrier
height, which freezes the spatial distribution of the atoms. One can then
count the number of atoms in the left or right well by measuring their
fluorescence. By implementing many of these measurements at different times,
the time evolution of the probabilities can be determined experimentally.

The initial state in our proposed scheme is realized through solving Eq. $%
\left( \ref{N21}\right) $ and Eq. $\left( \ref{N25}\right) $ with the
barrier height $d=300$ and plugging quasi-momentums $k_{1}$ and $k_{2}$
corresponding to the lowest energy into Eq. $\left( \ref{N2}\right) $. This
gives the initial state in the form $\left\vert \phi
_{0}(x_{1},x_{2})\right\rangle =\left( \left\vert LL\right\rangle
-\left\vert RR\right\rangle \right)/\sqrt{2} $, where $\left\vert
RR\right\rangle $ and $\left\vert LL\right\rangle $ denote that both atoms
reside in the ground state of the right or the left well. The correlation
function, also known as the two-body density, is defined as $\rho \left(
x_{1},x_{2},t\right) =\phi ^{\ast }(x_{1},x_{2},t)\phi (x_{1},x_{2},t)$.
Starting from the initial state $\left\vert
\phi_{0}(x_{1},x_{2})\right\rangle$, the time evolution of the wave function
is governed by $\left\vert \phi (x_{1},x_{2},t)\right\rangle =e^{-\frac{i}{%
\hbar}Ht}\left\vert \phi _{0}(x_{1},x_{2})\right\rangle =\sum_{i}C_{i}e^{-%
\frac{i}{\hbar}E_{i}t}\left\vert \Psi _{i}(x_{1},x_{2})\right\rangle $, in
which $C_{i}=\left\langle \Psi _{i}(x_{1},x_{2})|\phi
_{0}(x_{1},x_{2})\right\rangle $ is the overlap of the initial state and the 
$i$-th eigenstate of the system $\Psi _{i}(x_{1},x_{2})$ with eigen energy $%
E_{i}$ for barrier height $d=0.5$. The correlation function can be
calculated as 
\begin{eqnarray}
\rho \left( x_{1},x_{2},t\right) &=&\sum_{i=1}\left\vert C_{i}\Psi
_{i}(x_{1},x_{2})\right\vert ^{2}+\sum_{i<j}2C_{i}C_{j}  \notag \\
&&\times \Psi _{i}(x_{1},x_{2})\Psi _{j}(x_{1},x_{2})\cos \left[ \left(
E_{i}-E_{j}\right) t/\hbar \right] .  \notag \\
&&  \label{M2}
\end{eqnarray}

We introduce in theory the probabilities of finding $n$ particles in the
right or left well as $P_{Rn}\left( t\right) $ and $P_{Ln}\left( t\right) $.
The system remains population balanced at any subsequent time and it will be
as likely to find both atoms in the same well as in the opposite one, i.e. $%
P_{Rn}\left( t\right) =P_{Ln}\left( t\right) $. The mean particle number in
right well is $\overline{N}\left( t\right) =2P_{R2}\left( t\right)
+P_{R1}\left( t\right) $, which is a constant with the population-balanced
initial state and equals to $1$. Then the probability of finding both
particles in the same well is $P_{2}\left( t\right) =2P_{R2}\left( t\right)
=2\int_{0}^{L/2}\int_{0}^{L/2 }\rho \left( x_{1},x_{2},t\right) dx_{1}dx_{2}$
and that in different wells is $P_{1}\left( t\right) =P_{R1}\left( t\right)
=2\int_{-L/2 }^{0}\int_{0}^{L/2}\rho \left( x_{1},x_{2},t\right)
dx_{1}dx_{2} $ \cite{Zollner}. We emphasize that the pair tunneling between
double wells cancel each other in our system, while the single particle
tunneling will transfer the probabilities between double occupancy $%
P_{2}\left( t\right) $ and single occupancy $P_{1}\left( t\right) $.

\begin{figure}[tbp]
\includegraphics[width=0.50\textwidth]{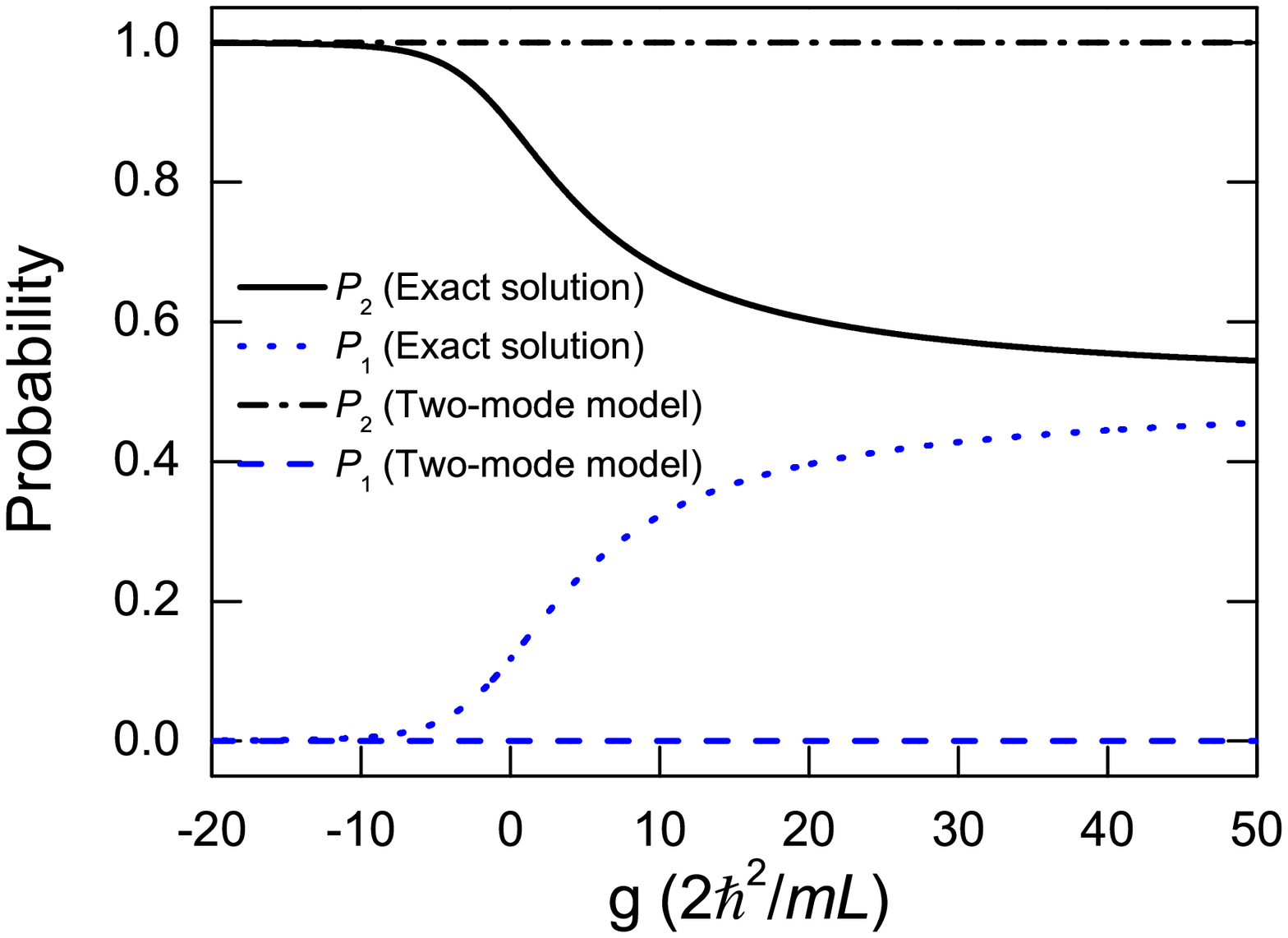} %
\includegraphics[width=0.50\textwidth]{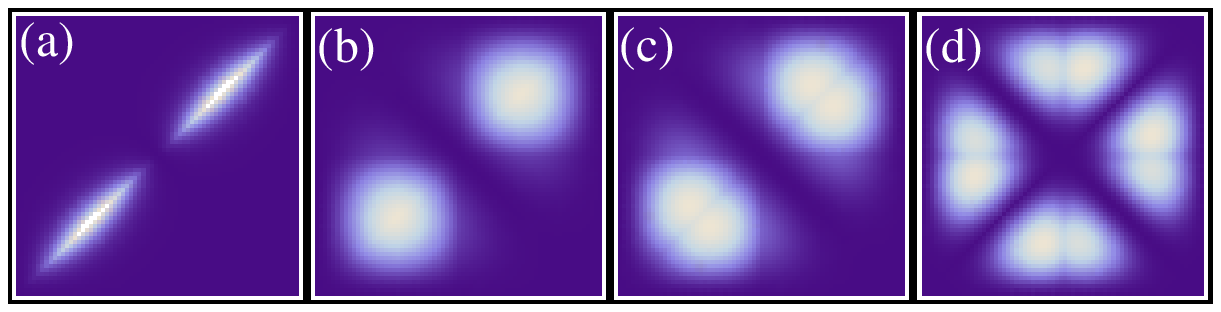}
\caption{(Color online). Top: Occupation probabilities as a function of the
interaction strength. The relative probabilities of both particles in the
same well $\left( P_{2}\text{, black curves}\right) $ or in different wells $%
\left( P_{1}\text{, blue curves}\right) $ are shown. The solid and dotted
lines show the results of our exact solution. The dash-dotted and dashed
curves show the those of two-mode model. Bottom: The two-body density $%
\protect\rho \left( x_{1},x_{2}\right) $ of the state $\left\vert
b\right\rangle $ for different interaction strengths (a) $g=-10$, (b) $g=-0.5
$, (c) $g=1.5$ and (d) $g=50$.}
\label{fig3}
\end{figure}

To better understand the oscillations in the probability, Fig. \ref{fig2}
presents the evolution of the two-body eigen energies $E_{i}$ as $g$ is
varied. In Fig. \ref{fig2}, we plot results of the spectrum for case $d=0.5$%
. Due to the odd symmetry of the initial state $\left\vert b\right\rangle$,
what we need in the time evolution are only those states with odd parity.
Note that the quasi-momentums of the bound states of atoms $\Psi(x_1,x_2)$
are real number, while those of the molecule states are two conjugate
complex numbers in the form of $k_{1,2}=\alpha \pm i\Lambda $, which only
happen for $g<0$. As we can see in the Eq. $\left( \ref{M2}\right) $, the
oscillation frequencies of the particles in double well are related to the
energy differences $\Delta E_{ij}=E_{i}-E_{j}$. Through calculating weight
coefficients $C_{i}$, we find that the first few $\Psi _{i}$'s make most
important contribution to the time evolution of correlation function $\rho
\left( x_{1},x_{2},t\right) $.

We propose here to experimentally realize the state $\left\vert
b\right\rangle $ and measure the influence of the interaction strength on
the distribution of the two particles between the wells. The occupation
probability of the eigenstate $\left\vert b\right\rangle =\Psi_1(x_1,x_2)$,
which is nothing but the lowest energy level in Fig. \ref{fig2}, is given in
Fig. \ref{fig3}. Together shown are the two-body density $\rho_1(x_1,x_2)$
for four typical values of $g$ with $d=0.5$. There exists a big difference
between the results of two-mode model and our exact solution. Both the
double occupancy and single occupancy probabilities of the state $\left\vert
b\right\rangle $ are constants in two-mode model. Clearly two-mode model is
not any more a good choice for double well, especially when the interaction
between two particles is strong or the excited states are involved in the
tunneling dynamics. Our calculation based on the exactly solved model
predict drastically different result: the double occupancy is suppressed for
increasing repulsive interactions, however, unlike in the case of ground
state $\left\vert a \right\rangle$, to an asymptotic value of 50\%. The
single occupancy probability, on the other hand, is enhanced to 50\% for
strong repulsion. The probabilities of double occupancy $P_2$ and single
occupancy $P_1$ tend towards equalization, which occurs in the experiment 
\cite{Murmann} for the ground state $\left\vert a\right\rangle $ and excited
state $\left\vert c\right\rangle $ of the non-interacting system. Attractive
interaction prefers the two atoms forming a molecule with large binding
energy, which will extinguish the probability of single particle tunneling
and double occupancy would dominate. The system approaches the two-particle
analog of a charge-density-wave state. The state $\left\vert b\right\rangle $
is the two-particle $NOON$ state, only when the attractive interaction is
large (an example is shown in Fig. \ref{fig3}($a$)) or the barrier height $d$
is infinity \cite{Liu}. The wave function of the odd parity symmetry state $%
\left\vert b\right\rangle $ is anti-symmetric about the nodal line $%
x_{1}=-x_{2}$. From the corresponding two-body density patterns shown in the
bottom of Fig \ref{fig3}, we see that for a fixed value of $d$ the density
for molecular states in attractive interaction has two peak values along the
diagonal line $x_{1}=x_{2}$. In the case of repulsive interaction the
probability accumulates gradually to the single occupancy region and finally
the probabilities $P_1$ and $P_2$ are equally distributed in the strong
interaction limit.

In order to explore the physical phenomenon behind the tunneling dynamics,
we investigate the effect of the interaction strength $g$ on oscillation of
the double occupancy probability $P_{2}\left( t\right) $ as shown in Fig. %
\ref{fig1}. We choose 30 eigenstates $\Psi_i(x_1,x_2)$ as the basis states
to study the tunneling dynamics, since the weight coefficients $C_{i}$ are
very small, when $i>30$. The time evolution of $P_{2}\left( t\right) $ is
illustrated in Fig. \ref{fig1} for repulsive interaction strengths from weak
to strong. The atoms tunnel back and forth between the double well, as a
result the probabilities $P_2$ oscillate with time and the amplitude
increases with $g$ due to the enhancement of the tunneling rate. Stronger
repulsion will destroy the perfect oscillation such that the double
occupancy probability in the initial $NOON$ state can not be retrieved with
100\% fidelity. The oscillation period shrinks from $\tau=0.64$ for $g=1.5$
to $\tau=0.6$ for $g=20$. The two-body density in the bottom of Fig. \ref%
{fig1} shows that the probability will be redistributed in blocks $n=0,2$
and $1$, leaving partial residual probability after the completion of an
oscillation.

Experimental realization of our $\delta$-split double well includes
producing the quasi-1D waveguide and loading two atoms in the motional
ground state of a single optical microtrap. A cylindrically focused
blue-detuned Gaussian laser beam directed perpendicular to the long axis of
the confining potential is used to cut the waveguide in half \cite{Nguyen}. A series of
such laser beams with equal spatial distance create a lattice model.
Moreover, the ability of preparing the two atoms in the excited states $%
\left\vert b\right\rangle $ and $\left\vert c\right\rangle $ allows for a
direct realization of population of higher bands, which contains the physics
responsible for the formation of novel ordered phases in many-body system 
\cite{Muller,Wirth}.

\textit{Conclusions:} We present the Bethe-ansatz type exact solution for $N$
interacting bosonic atoms in the $\delta $-split double well, the result of
which for two atoms are applied to the tunneling dynamics starting from an
odd parity state $\left\vert b\right\rangle $. The occupation probability
show evident dependence on the interaction and tend towards $50\%-50\%$
equalization which is different from the result of two-mode model adopted in
the experiment. The probabilities are found to oscillate between double and
single occupancy in the tunneling dynamics.

\begin{acknowledgments}
This work is supported by NSF of China under Grant Nos. 11234008 and
11474189, the National Basic Research Program of China (973 Program) under
Grant No. 2011CB921601, Program for Changjiang Scholars and Innovative
Research Team in University (PCSIRT)(No. IRT13076).
\end{acknowledgments}

\end{document}


\title{Supplementary Material of Two Atoms in a Double Well: an Exact
Solution}
\author{Yanxia Liu}
\affiliation{Institute of Theoretical Physics, Shanxi University, Taiyuan 030006, P. R.
China}
\author{Yunbo Zhang}
\email{ybzhang@sxu.edu.cn}
\affiliation{Institute of Theoretical Physics, Shanxi University, Taiyuan 030006, P. R.
China}
\maketitle

\appendix

\section{Bethe-ansatz-type equations}

\begin{figure}[tbp]
\includegraphics[width=0.45\textwidth]{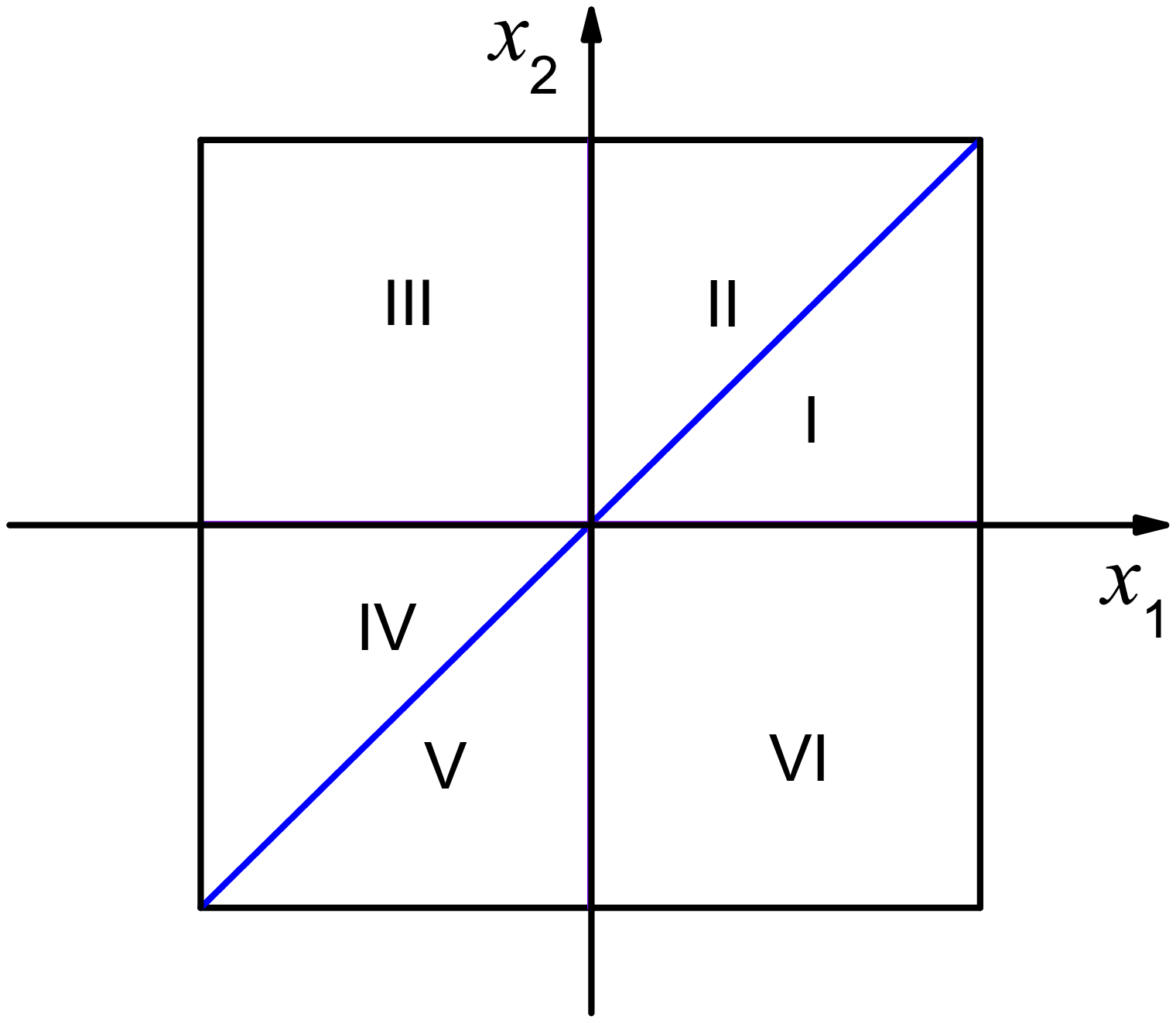}
\caption{(Color online). The square well with infinite depth can be
dissolved into six regions by the $x_{1}$ axis, $x_{2}$ axis, and the line $%
x_{1}=x_{2}$. The adjacent regions are connected by the continuous
conditions of wave functions and jump conditions of the first derivative of
wave functions. Away from these lines, the particles are free.}
\label{fig1}
\end{figure}

Here we show the details on how to exactly solve the system of interacting $%
N $ particles in the 1D $\delta$-barrier split double well. For this system,
the key to the exact solution is region allocation. Denote the configuration
of $n$ particles in the left well as block $n$. We connect the adjacent
regions in a fixed block $n$, and then connect the blocks $n$ and $n-1$ step
by step. Take the two particles case as an example. We can transform the
two-particle 1D problem into a 2D problem of one particle, as shown in Fig. %
\ref{fig1}. We divide the $x_1x_2$ box into six regions. The regions I and
II constitute block $0$, region III and VI constitute block $1$ and IV and V
constitute block $2$. Detailed analysis procedure of the $N$ particle cases
is demonstrated below. 
For bosons the wave function of the system is symmetric under coordinates
exchange%
\begin{equation}
\Psi (\cdots x_{i},\cdots x_{j}\cdots )=\Psi (\cdots x_{j},\cdots
x_{i}\cdots ),
\end{equation}%
which means%
\begin{equation}
A_{n}\left( Q,\epsilon _{P}P\right) =A_{n}\left( Q^{\prime },\epsilon
_{P}P\right) .
\end{equation}%
Thus for fixed $n$ the coefficients are equal for all local regions assigned
by arrangements $Q^{\prime }=\left( Q_{1}^{\prime },Q_{2}^{\prime
},Q_{3}^{\prime },\cdots ,Q_{N}^{\prime }\right) $. In this block, we
noticed that the open boundary conditions are imposed on the left end of the
tube 
\begin{equation}
\Psi \left( x_{Q_{i}}=-\frac{L}{2}\right) =0,i=1,2,\cdots n,  \label{Left}
\end{equation}%
for particles in the left side and on the right end 
\begin{equation}
\Psi \left( x_{Q_{i}}=\frac{L}{2}\right) =0,i=n+1,\cdots N.  \label{Right}
\end{equation}%
for particles in the right well. We define the reflection matrix $R\left(
P_{1}\right) $%
\begin{equation}
R\left( P_{1}\right) =\frac{A_{n}\left( Q,-P_{1},\cdots ,\epsilon
_{P_{N}}P_{N}\right) }{A_{n}\left( Q,P_{1},\cdots ,\epsilon
_{P_{N}}P_{N}\right) }=-\exp \left( -ik_{P_{1}}L\right) ,  \label{p5}
\end{equation}%
and $R\left( P_{N}\right) $%
\begin{equation}
R\left( P_{N}\right) =\frac{A_{n}\left( Q,\epsilon _{P_{1}}P_{1},\cdots
,-P_{N}\right) }{A_{n}\left( Q,\epsilon _{P_{1}}P_{1},\cdots ,P_{N}\right) }%
=-\exp \left( ik_{P_{N}}L\right) .  \label{p4}
\end{equation}%
Two special cases are that all particles are located on the right or left
side of the barrier such that we have only half of the boundary condition (%
\ref{Right}) or (\ref{Left}). Thus only the reflection matrix $R\left(
P_{N}\right) $ or $R\left( P_{1}\right) $ survives for block $n=0$ or $N$,
respectively, and only the rightmost or leftmost atom can be reflected on
the boundaries and get a phase shift.

Next we discuss the scattering between two particles. Let $x_{i}$ and $x_{j}$
be coordinates of two adjacent particles in the same side of the barrier.
Particles in different well cannot collide with each other. In the relative
coordinates the first derivative of wave function is not continuous due to
the $g$-interaction, 
\begin{equation*}
\left[ \left( \frac{\partial \Psi }{\partial x_{i}}-\frac{\partial \Psi }{%
\partial x_{j}}\right) |_{x_{i}-x_{j}=0}\right] -\frac{m}{\hbar ^{2}}g\Psi
|_{x_{i}-x_{j}=0}=0,
\end{equation*}%
which connects different local regions in one block. Generally there are $%
N\left( N-1\right) /2$ such equations in each block and they are equivalent
to each other for different blocks. One exceptional case is $N=2$: we have
one equation for blocks $n=0$ or $2$, while no $g$-interaction condition in
block $n=1$. In a similar way we define the scattering matrix $S_{\epsilon
_{P_{i}}P_{i},\epsilon _{P_{j}}P_{j}}$ describing the elastic collision
between the two particles 
\begin{eqnarray}
S_{\epsilon _{P_{i}}P_{i},\epsilon _{P_{j}}P_{j}} &=&\frac{A_{n}\left(
Q,\cdots ,\epsilon _{P_{j}}P_{j},\epsilon _{P_{i}}P_{i},\cdots \right) }{%
A_{n}\left( Q,\cdots ,\epsilon _{P_{i}}P_{i},\epsilon _{P_{j}}P_{j},\cdots
\right) }  \notag \\
&=&\frac{\epsilon _{P_{i}}k_{P_{i}}-\epsilon _{P_{j}}k_{P_{j}}-i\frac{m}{%
\hbar ^{2}}g}{\epsilon _{P_{i}}k_{P_{i}}-\epsilon _{P_{j}}k_{P_{j}}+i\frac{m%
}{\hbar ^{2}}g}.  \label{p7}
\end{eqnarray}%
The matrices $R\left( P_{1}\right) $, $R\left( P_{N}\right) $ and $%
S_{\epsilon _{P_{i}}P_{i},\epsilon _{P_{j}}P_{j}}$ are defined in the same
way as in Lieb-Liniger model with hard wall boundary conditions. The missing
reflection and scattering matrices make it impossible to get the Bethe
ansatz equation in a single block $n$. We shall see in the following, the
jump conditions between blocks with different $n$ play essential role in the
exact solution of our model.

The jump condition due to the $d$-barrier at $x_{Q_{n}}=0$ connects the
blocks $n-1$ and $n$, which can be obtained by integrating the Schr\"{o}%
dinger equation from $x_{Q_{n}}=0_{-}$ to $x_{Q_{n}}=0_{+}$ 
\begin{equation}
\left[ \frac{\partial \Psi }{\partial x_{Q_{n}}}|_{x_{Q_{n}}=0_{+}}-\frac{%
\partial \Psi }{\partial x_{Q_{n}}}|_{x_{Q_{n}}=0_{-}}\right] -\frac{2m}{%
\hbar ^{2}}d\Psi |_{x_{Q_{n}}=0}=0.\text{ }  \label{jumpd}
\end{equation}%
It means that the $n$-th particle jumps from the right to the left.
Inserting the Bethe-ansatz wave function Eq. (2) into Eq. (\ref{jumpd}), we
get the relation 
\begin{eqnarray}
&&\left( ik_{P_{n}}-\frac{m}{\hbar ^{2}}d\right) A_{n-1}\left( Q,\cdots
,P_{n},\cdots \right)   \notag \\
&&-\left( ik_{P_{n}}+\frac{m}{\hbar ^{2}}d\right) A_{n-1}\left( Q,\cdots
,-P_{n},\cdots \right)   \notag \\
&&-\left( ik_{P_{n}}+\frac{m}{\hbar ^{2}}d\right) A_{n}\left( Q,\cdots
,P_{n},\cdots \right)   \notag \\
&&+\left( ik_{P_{n}}-\frac{m}{\hbar ^{2}}d\right) A_{n}\left( Q,\cdots
,-P_{n},\cdots \right) =0.  \label{p16}
\end{eqnarray}%
Seemingly cumbersome, this is a basic relationship between four coefficients 
$A$ in blocks $n-1$ and $n$. Through the above analysis, we know that to
reverse the sign of the $n$-th particle's momentum in block $n-1$, the
particle needs firstly to scatter with the right neighboring atoms for $N-n$
times until there are no more atoms to scatter, then reflect on the boundary
and finally scatter with the left neighboring atoms for the same times. This
successive scattering process will reverse the sign of $P_{n}$ and by
repeatedly using the matrices $S_{\epsilon _{P_{i}}P_{i},\epsilon
_{P_{j}}P_{j}}$ and $R(P_{N})$ we can get the Eq. $\left( 4\right) $. Note
that $R_{n-1}(P_{n})=R(P_{N})$ for $n=N$. In a similar way we can inverse
the momentum of the $n$-th particle in the block $n$ by moving first toward
left, reflecting on the boundary $-L/2$ and going back to its original
position toward right. This gives the Eq. $\left( 5\right) $. Similarly we
have $R_{n}(P_{n})=R(P_{1})$ for $n=1$. In Eq. (\ref{p16}) we replace the
coefficients $A$ with $-P_{n}$ by their counterpart with $P_{n}$ through Eq.
($4$) and Eq. ($5$) and the scattering matrix between the $d$-barrier and
the $n$-th particle is calculated as 
\begin{eqnarray}
S_{d,P_{n}}\left( n\right)  &=&\frac{A_{n}\left( Q,\cdots ,P_{n},\cdots
\right) }{A_{n-1}\left( Q,\cdots ,P_{n},\cdots \right) }  \notag  \label{N19}
\\
&=&\frac{ik_{P_{n}}\left( 1-R_{n-1}\left( P_{n}\right) \right) -\frac{m}{%
\hbar ^{2}}d\left( 1+R_{n-1}\left( P_{n}\right) \right) }{ik_{P_{n}}\left(
1-R_{n}\left( P_{n}\right) \right) +\frac{m}{\hbar ^{2}}d\left(
1+R_{n}\left( P_{n}\right) \right) },  \notag \\
&&
\end{eqnarray}%
which brings the system from block $n-1$ to $n$. The continuous condition of
wave function $\Psi |_{x_{Q_{n}}=0_{+}}=\Psi |_{x_{Q_{n}}=0_{-}}$, on the
other hand, gives an alternative form of this scattering matrix with $d$%
-barrier, that is, Eq. $\left( 7\right) $. Then we infer that in presence of
the potential barrier the momentum set $\{k\}$ satisfy the BA-type equations
(3). For $N=2$, the BA-type equation can be written as 
\begin{equation}
ik_{1}\left( \frac{1+\frac{k_{2}-k_{1}+i\frac{m}{\hbar ^{2}}g}{k_{2}-k_{1}-i%
\frac{2\mu }{\hbar ^{2}}g}\cdot \frac{k_{2}+k_{1}-i\frac{m}{\hbar ^{2}}g}{%
k_{2}+k_{1}+i\frac{m}{\hbar ^{2}}g}\exp \left( ik_{1}L\right) }{1-\frac{%
k_{2}-k_{1}+i\frac{m}{\hbar ^{2}}g}{k_{2}-k_{1}-i\frac{2\mu }{\hbar ^{2}}g}%
\cdot \frac{k_{2}+k_{1}-i\frac{m}{\hbar ^{2}}g}{k_{2}+k_{1}+i\frac{m}{\hbar
^{2}}g}\exp \left( ik_{1}L\right) }-\frac{1+\exp \left( -ik_{1}L\right) }{%
1-\exp \left( -ik_{1}L\right) }\right) =\frac{2m}{\hbar ^{2}}d.  \label{p10}
\end{equation}%
Another BA-type equation can be obtained by swapping the positions of $k_{1}$ and $%
k_{2}$.
\begin{figure}[tbp]
\includegraphics[width=0.50\textwidth]{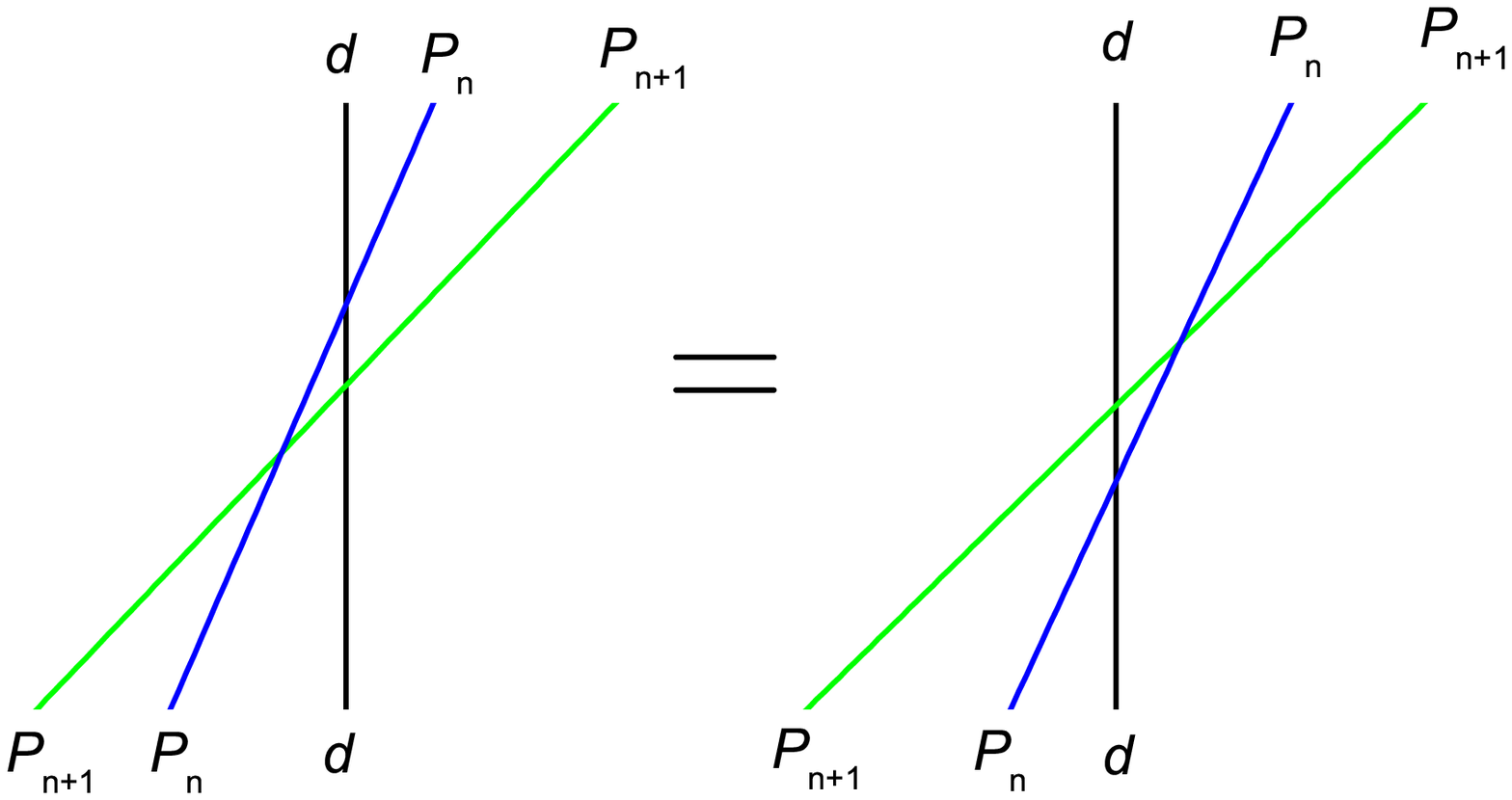}
\caption{(Color online). Graphical representation of two possible ways of
two particles scattering with $d$-barrier.}
\label{fig2}
\end{figure}

\begin{figure}[tbp]
\includegraphics[width=0.45\textwidth]{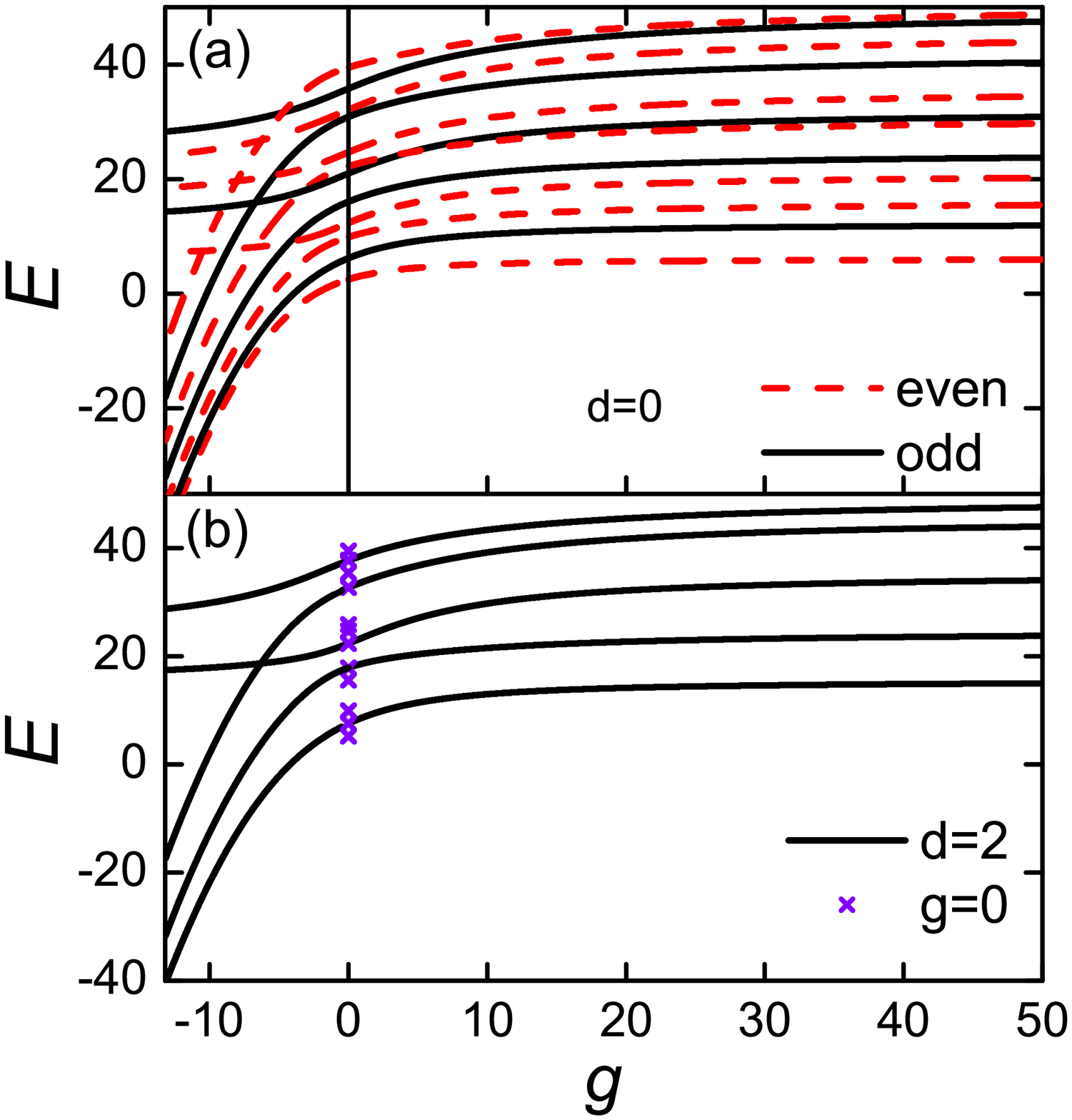}
\caption{(Color online). The energy spectrum for the system of particles in
1D square well with infinite depth (a) $d=0$ and particles in
1D split hard-wall potential (b) $d=2$. The $\times $ signs indicate the 
energy for $g=0$. The (red) dash lines show the spectrum for
even parity states. The (black) solid lines show the spectrum for odd parity states.}
\label{fig3}
\end{figure}

\begin{figure}[tbp]
\includegraphics[width=0.5\textwidth]{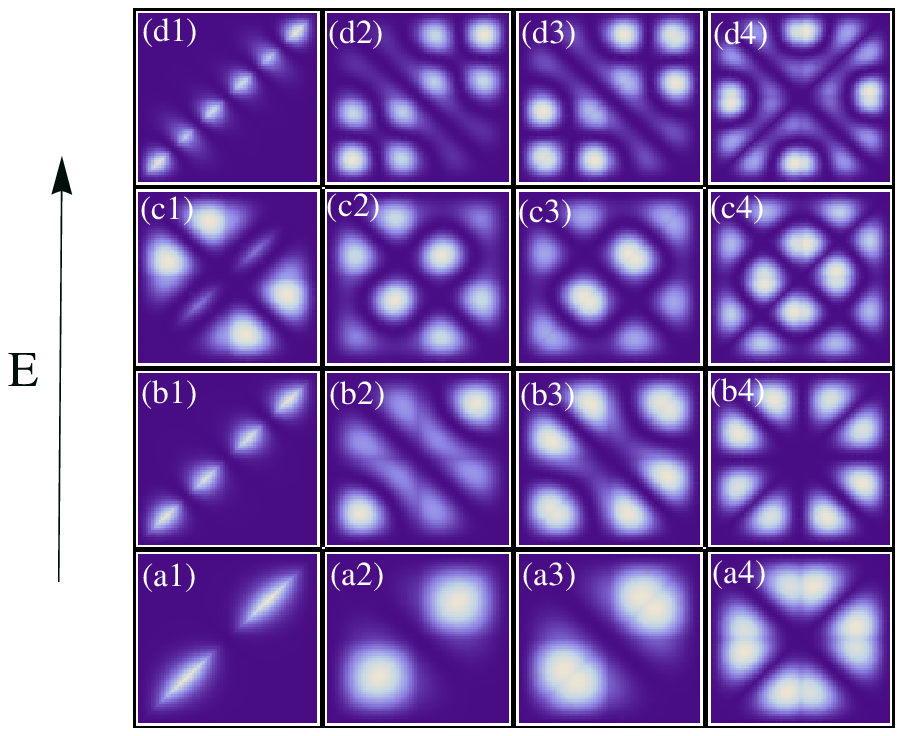}
\caption{(Color online). The normalized probability density $\protect\rho %
\left( x_{1},x_{2}\right) $ for two particles in a double-well potential with  
$d=2$ for different $g$. The columns represent results for four interaction parameters 
$g=-6.295, -0.5, 1.5, 50$, respectively. From bottom to up the rows are for four lowest states.
Each individual plot spans the range $-1<x_{1},x_{2}<1$.}
\label{fig4}
\end{figure}

\section{Yang-Baxter-Type relation}

Since Yang and Baxter's pioneering works, the quantum Yang-Baxter equation
(QYBE), which define the underlying algebraic structure, has become a
cornerstone for constructing and solving the integrable models.
Yang-Baxter-Type relation ($6$) can be represented graphically. In Fig. \ref%
{fig2} we show two equivalent scattering ways of two particles with the
barrier. The black vertical line indicates that the barrier is localized in
the center. Each intersection between two lines is represented by a
scattering matrix. The arrow of time is taken from top to bottom. There are
two distinct forms of S-matrix, that between particles and that between the
particle and the barrier. For $N=2$, Eq. (8) can be written as 
\begin{eqnarray}
&&\sin ^{2}\left( \frac{k_{1}L}{2}\right) \sin ^{2}\left( -\arctan \left( 
\frac{\frac{m}{\hbar ^{2}}g}{k_{1}+k_{2}}\right) +\arctan \left( \frac{\frac{%
m}{\hbar ^{2}}g}{k_{1}-k_{2}}\right) +\frac{k_{2}L}{2}\right)  \notag \\
&=&\sin ^{2}\left( \frac{k_{2}L}{2}\right) \sin ^{2}\left( -\arctan \left( 
\frac{\frac{m}{\hbar ^{2}}g}{k_{1}+k_{2}}\right) -\arctan \left( \frac{\frac{%
m}{\hbar ^{2}}g}{k_{1}-k_{2}}\right) +\frac{k_{1}L}{2}\right)  \label{p9}
\end{eqnarray}

\section{The energy spectrum for two particles in a double well}

The energy spectrum for two particles in a double well is obtained by
numerically solving the three transcendental equations Eq. (\ref{p10}) and
Eq. (\ref{p9}) for given barrier height $d$. In Fig. \ref{fig3}, we plot the
energy spectrum as a function of $g$ for limiting case $d=0$ and a moderate
barrier height $d=2$. In the absence of $\delta $-function potential barrier
we find two kinds of parity symmetry bound states [see Fig. \ref{fig3}$%
\left( a\right) $], while there are only odd parity bound states for finite
potential barrier [Fig. \ref{fig3}$\left( b\right) $] except at $g=0$. Note
a level crossing between the bound state of atoms and the molecular state
occurs at $g=-6.295$.

The two-body density $\rho (x_{1}, x_{2})$ for a fixed barrier height $d=2$
is presented in Fig. \ref{fig4}. From the ground state density in first
row of Fig. \ref{fig4} (a1-a4) we see almost no difference with a lower
barrier $d=0.5$. Again as in Fig. 4 in the main text we choose four typical
interaction strength parameters $g=-6.295, -0.5, 1.5, 50$, one of which is
exactly the level crossing point. There we see that for a fixed value of $d$, 
the densities for molecular states reside along the diagonal line $x_{1}=x_{2}$.
For repulsive interactions, the probability of double occupancy is transferred
into the single occupancy region as $g$ increases and the two finally reach
equalization for strong interaction. The two-body densities for higher excited 
states in Fig. \ref{fig4} (b-d) show more bizarre oscillations and more nodal lines
divided the plane into small probability islands, unveiling 
an equal distribution of the double and single occupancy probability 
for strong repulsive interaction. We also note the big difference for the 
molecular state and the atomic state. While there are always probability
for the single occupancy in the atomic states, the molecular state tend to 
exclude it. An typical example is shown in panels (c1) and (d1).
 
\begin{figure}[tbp]
\includegraphics[width=0.50\textwidth]{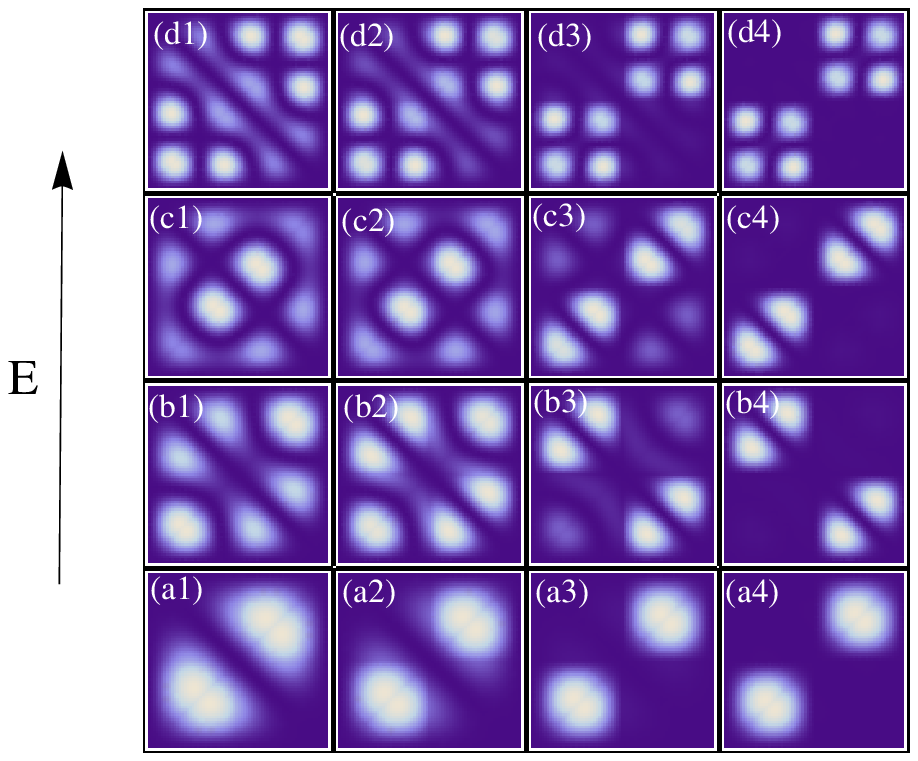}
\caption{(Color online). The normalized probability density $\protect\rho %
\left( x_{1},x_{2}\right) $ for two particles in a double-well potential for 
$g=1.5$ with different $d$. The columns represent results for four barrier height 
$d=0.5, 1.5, 10, 50$, respectively. From bottom to up the rows are for four lowest states. 
Each individual plot spans the range $-1<x_{1},x_{2}<1$.}
\label{fig5}
\end{figure}

Fig. \ref{fig5} shows the two-body density $\rho (x_{1}, x_{2})$ with a fixed value
$g=1.5$ for different $d$. In the lowest odd symmetry state the density tends to 
converge towards the first and third quadrants and the probability of double 
occupancy $P_{2}$ is approaching $1$ as $d$ increases, which eventually achieves  
the $NOON$ state in (a4). While we observe similar tendency for the third and the fourth
states, the density of the some higher excited states (the second
one in Fig. \ref{fig5}(b1-b4), also the fifth one which is not shown) exhibits opposite behavior, 
which tends to converge towards the second and fourth quadrants and probability $P_{2}$ 
is approaching $0$ for higher barrier. The excited states are classified according to the 
total energy of both atoms proportional to $k_1^2+k_2^2$. In extreme case of $g=0$ with
very high barrier, (a4) and (d4) reduce to the states with almost equal quasi-momentum $k$, 
and the energy for (b4) and (c4) is nearly degenerate.

\section{Preparation of the initial state}

The system is initially prepared in the odd parity state with barrier height $d=300$, the 
wave function of which can be obtained through solving Eq. (\ref{p10}) and Eq. (\ref%
{p9}) for different interaction strength. The numerically computed quasi-momentums 
are $k_{1}=2.734$ and $k_{2}=3.6936$ for $g=0.5$, $k_{1}=4.126$ and $k_{2}=2.585$ for 
$g=1.5$, and $k_{1}=2.872$ and $k_{2}=5.726$ for $g=20$.
These $k_{1}$ and $k_{2}$ are plugged into Eq. $\left(2\right) $ to get the corresponding 
wave function for the initial state, which can be regarded as the $NOON$ state. 
Then we lower the barrier height from $d=300 $ to $d=0.5$ and let the system to evolve
with time. Therefore we choose the eigenstates with barrier height $d=0.5$
as a complete set, upon which the initial state is expanded. Here we only choose 30 eigenstates, since the
weight coefficients $C_{i}$ are very small when $i>30$, which are shown in
the Table. \ref{table1}

\begin{table}[tbp]
\caption{The weight coefficients $C_{i}$}
\label{table1}
\begin{center}
\centering%
\begin{tabular}{cccccccc}
\hline
$C_{i}$ & $g=0.5$ & $g=1.5$ & $g=20$ & $\ \ \ \ C_{i}$ & $g=0.5$ & $g=1.5$ & 
$g=20$ \\ \hline
$C_{1}$ & $-0.8731618$ & $0.8524551$ & $0.666279$ & $\ \ \ \ C_{16}$ & $%
0.0000835$ & $0.0001781$ & $-0.00021$ \\ 
$C_{2}$ & $0.46679$ & $-0.499186$ & $-0.66614$ & $\ \ \ \ C_{17}$ & $%
-0.0328012$ & $0.0338401$ & $0.024499$ \\ 
$C_{3}$ & $-0.0188581$ & $0.0534258$ & $-0.24749$ & $\ \ \ \ C_{18}$ & $%
0.0007173$ & $0.0021096$ & $-0.00839$ \\ 
$C_{4}$ & $0.010067$ & $0.0292036$ & $-0.1765$ & $\ \ \ \ C_{19}$ & $%
-0.0001443$ & $-0.000379$ & $-0.00032$ \\ 
$C_{5}$ & $-0.1179361$ & $0.1204126$ & $0.065573$ & $\ \ \ \ C_{20}$ & $%
-0.0000502$ & $-0.000114$ & $-0.000057$ \\ 
$C_{6}$ & $-0.0024673$ & $-0.0066555$ & $-0.01626$ & $\ \ \ \ C_{21}$ & $%
-0.0007583$ & $0.0022915$ & $0.0193$ \\ 
$C_{7}$ & $0.0040574$ & $0.0119879$ & $0.069451$ & $\ \ \ \ C_{22}$ & $%
-0.0004255$ & $-0.0014146$ & $-0.01899$ \\ 
$C_{8}$ & $0.0022943$ & $0.0074603$ & $0.061807$ & $\ \ \ \ C_{23}$ & $%
0.0000968$ & $0.0002537$ & $-0.00169$ \\ 
$C_{9}$ & $0.000471$ & $0.0010114$ & $-0.00419$ & $\ \ \ \ C_{24}$ & $%
0.0000447$ & $0.0001115$ & $-0.00017$ \\ 
$C_{10}$ & $0.0558227$ & $0.0573916$ & $0.036043$ & $\ \ \ \ C_{25}$ & $%
0.0000239$ & $0.0000509$ & $-0.000025$ \\ 
$C_{11}$ & $0.0012207$ & $-0.0035508$ & $-0.01138$ & $\ \ \ \ C_{26}$ & $%
0.0216447$ & $-0.02237$ & $-0.01812$ \\ 
$C_{12}$ & $-0.0002296$ & $0.000539$ & $0.000308$ & $\ \ \ \ C_{27}$ & $%
-0.0004727$ & $-0.0013925$ & $-0.00643$ \\ 
$C_{13}$ & $-0.0015487$ & $-0.0046468$ & $-0.03337$ & $\ \ \ \ C_{28}$ & $%
0.0000958$ & $0.0002573$ & $0.000285$ \\ 
$C_{14}$ & $-0.0008721$ & $0.0028878$ & $0.031884$ & $\ \ \ \ C_{29}$ & $%
-0.0000357$ & $-0.0000915$ & $-0.000068$ \\ 
$C_{15}$ & $0.0001965$ & $0.0004967$ & $-0.00269$ & $\ \ \ \ C_{30}$ & $%
-0.0000161$ & $-0.0000359$ & $-0.000017$ \\ \hline
\end{tabular}%
\end{center}
\end{table}